\documentclass[conference]{IEEEtran}
\IEEEoverridecommandlockouts
\usepackage{cite}
\usepackage{amsmath,amssymb,amsfonts}
\usepackage{algorithmic}
\usepackage{graphicx}
\usepackage{textcomp}
\usepackage{xcolor}
\usepackage{balance}
\usepackage{algorithm}
\usepackage{algorithmic}
\usepackage{amsmath}% for double integral symbol in this template
\usepackage{booktabs}
\usepackage{amsfonts}
\usepackage{microtype}  
\usepackage[list=true]{subcaption}
\usepackage[font={footnotesize}]{caption}
\usepackage{times}% use times font for the paper instead of default Computer Modern fonts
\usepackage{multirow}% to allow multiple-row elements in tabular environment
\usepackage[none]{hyphenat}% turn off hyphenation to make text extraction and indexing easier
\usepackage{float}% better control of floating figures and tables
\usepackage{acro}
\usepackage{tikz}
\usepackage{pgfplots}
\usetikzlibrary{svg.path}
\tikzset{every picture/.style={line width=0.6pt}}
\usetikzlibrary{calc,positioning}
\usetikzlibrary{arrows.meta,
                backgrounds,
                chains,
                fit,
                quotes}
\usepackage[bottom]{footmisc}

\DeclareAcronym{td}{
short=TD,
long= time domain,
}
\DeclareAcronym{fd}{
short=FD,
long= frequency domain,
}

\DeclareAcronym{5g}{
short=5G,
long= fifth generation,
}
\DeclareAcronym{mu}{
short=MU,
long= multi-user,
}
\DeclareAcronym{cnn}{
short=CNN,
long= convolutional neural network,
}
\DeclareAcronym{csi}{
short=CSI,
long= channel state information,
}
\DeclareAcronym{zf}{
short=ZF,
long= zero-forcing,
}
\DeclareAcronym{ici}{
short=ICI,
long= intercarrier interference,
}
\DeclareAcronym{dft}{
short=DFT,
long= discrete Fourier transform,
}
\DeclareAcronym{2d}{
short=2D,
long= two-dimensional,
}
\DeclareAcronym{idft}{
short=IDFT,
long= inverse discrete Fourier transform,
}
\DeclareAcronym{bs}{
short=BS,
long= base station,
}
\DeclareAcronym{ue}{
short=UE,
long= user equipment,
}
\DeclareAcronym{iq}{
short=I/Q,
long= quadrature,
}
\DeclareAcronym{I}{
short=I,
long= in-phase,
}
\DeclareAcronym{Q}{
short=Q,
long= quadrature,
}
\DeclareAcronym{ls}{
short=LS,
long= least squares,
}
\DeclareAcronym{ota}{
short=OTA,
long= over-the-air,
}
\DeclareAcronym{sgd}{
short=SGD,
long= stochastic gradient descent,
}
\DeclareAcronym{ce}{
short=CE,
long= cross-entropy,
}
\DeclareAcronym{sl}{
short=SL,
long= supervised learning,
}
\DeclareAcronym{rl}{
short=RL,
long= reinforcement learning,
}
\DeclareAcronym{awgn}{
short=AWGN,
long= additive white Gaussian noise,
}
\DeclareAcronym{ser}{
short=SER,
long= symbol error rate,
}
\DeclareAcronym{qam}{
short=QAM,
long= quadrature amplitude modulation,
}
\DeclareAcronym{rrc}{
short=RRC,
long= root-raised cosine,
}

\DeclareAcronym{snr}{
short=SNR,
long= signal-to-noise ratio,
}
\DeclareAcronym{rvftdnn}{
short=RVFTDNN,
long= real-valued focused time-delay neural network,
}
\DeclareAcronym{lo}{
short=LO,
long= local oscillator,
}
\DeclareAcronym{lpf}{
short=LPF,
long= lowpass filter,
}
\DeclareAcronym{pdf}{
short=PDF,
long= probability density function,
}
\DeclareAcronym{cdf}{
short=CDF,
long= cumulative distribution function ,
}
\DeclareAcronym{fir}{
short=FIR,
long= finite impulse response,
}
\DeclareAcronym{rhs}{
short=RHS,
long= right-hand side,
}
\DeclareAcronym{dsp}{
short=DSP,
long= digital signal processing,
}
\DeclareAcronym{nn}{
short=NN,
long= neural network,
}
\DeclareAcronym{mlp}{
short=MLP,
long=multilayer perceptron
}
\DeclareAcronym{GaN}{
short=GaN,
long=Gallium Nitride,
}
\DeclareAcronym{relu}{
short=ReLU,
long = rectified linear unit, 
}
\DeclareAcronym{mse}{
short=MSE,
long=mean squared error,
}
\DeclareAcronym{rvtdnn}{
short=RVTDNN,
long= real-valued time-delay neural network,
}
\DeclareAcronym{arvtdnn}{
short=ARVTDNN,
long= augmented real-valued time-delay neural network,
}
\DeclareAcronym{arden}{
short=ARDEN,
long= attention residual real-valued time-delay neural network,
}
\DeclareAcronym{r2tdnn}{
short=R2TDNN,
long= residual real-valued time-delay neural network,
}
\DeclareAcronym{flop}{
short=FLOP,
long= floating point operations,
}
\DeclareAcronym{ph}{
short=PH,
long= parallel Hammerstein
}
\DeclareAcronym{ofdm}{
short=OFDM,
long=orthogonal frequency division multiplexing,
}
\DeclareAcronym{par}{
short=PAR,
long=peak-to-average ratio,
}
\DeclareAcronym{papr}{
short=PAPR,
long=peak-to-average power ratio,
}
\DeclareAcronym{rf}{
short=RF,
long=radio frequency,
}
\DeclareAcronym{pa}{
short=PA,
long=power amplifier,
}
\DeclareAcronym{pas}{
short=\acs{pa}s,
long=power amplifiers,
}
\DeclareAcronym{psd}{
short=PSD,
long= power spectral density,
}
\DeclareAcronym{dpd}{
short=DPD,
long=digital predistortion,
}
\DeclareAcronym{cfr}{
short=CFR,
long=crest factor reduction,
}
\DeclareAcronym{cf}{
short=CF,
long=crest-factor}

\DeclareAcronym{evm}{
short=EVM,
long=error vector magnitude,
}
\DeclareAcronym{nmse}{
short=NMSE,
long=normalized mean squared error,
}
\DeclareAcronym{acpr}{
short=ACPR,
long=adjacent channel power ratio,
}
\DeclareAcronym{pae}{
short=PAE,
long=power added efficiency,
}
\DeclareAcronym{dla}{
short=DLA,
long=direct learning architecture,
}
\DeclareAcronym{ila}{
short=ILA,
long=indirect learning architecture,
}
\DeclareAcronym{ilc}{
short=ILC,
long=iterative learning control ,
}
\DeclareAcronym{cfr-dpd}{
short=CFR-DPD,
long=CFR combined with DPD,
}
\DeclareAcronym{icf}{
short=ICF,
long=iterative clipping and filtering,
}
\DeclareAcronym{am/am}{
short=AM/AM,
long=amplitude-to-amplitude,
}
\DeclareAcronym{am/pm}{
short=AM/PM,
long=amplitude-to-phase,
}
\DeclareAcronym{siso}{
short=SISO,
long=single-input single-output
}
\DeclareAcronym{mimo}{
short=MIMO,
long=multiple-input multiple-output
}
\DeclareAcronym{mp}{
short=MP,
long=memory polynomial
}
\DeclareAcronym{gmp}{
short=GMP,
long=generalized memory polynomial
}
\DeclareAcronym{adc}{
short=ADC,
long= analog-to-digital converter}
\DeclareAcronym{dac}{
short=DAC,
long= digital-to-analog converter}
\DeclareAcronym{ilc-dpd}{
short=ILC-DPD,
long= adaptive ILC-based DPD
}
\DeclareAcronym{rms}{
short=RMS,
long= root mean squares
}
\DeclareAcronym{vst}{
short=VST,
long= vector signal transceiver
}
\DeclareAcronym{mmwv}{
short=mm-Wave,
long= millimeter-wave
}

\begin{document}
\bstctlcite{IEEEexample:BSTcontrol}

\title{Frequency-domain digital predistortion \\for Massive MU-MIMO-OFDM Downlink\\
\thanks{This work was supported by the Swedish Foundation for Strategic Research (SSF), grant no. ID19-0021. The authors would like to thank Fan Jiang at Chalmers University of Technology for fruitful discussions.}}

\author{Yibo~Wu\IEEEauthorrefmark{1}\IEEEauthorrefmark{2},
        Ulf~Gustavsson\IEEEauthorrefmark{1},
      Mikko~Valkama\IEEEauthorrefmark{3},        Alexandre~Graell~i~Amat\IEEEauthorrefmark{2}, and
        Henk~Wymeersch\IEEEauthorrefmark{2}\\
        \IEEEauthorrefmark{1}Ericsson Research, Gothenburg, Sweden\\
        \IEEEauthorrefmark{2}Department of Electrical Engineeering, Chalmers University of Technology, Gothenburg, Sweden\\
		\IEEEauthorrefmark{3}Department of Electrical Engineering, Tampere University,  Tampere, Finland
        % \\email: yibo@chalmers.se
        }

\maketitle

\begin{abstract}
\Acf{dpd} is a method commonly used to compensate for the nonlinear effects of  \acp{pa}. However, the computational complexity  of most \ac{dpd} algorithms becomes an issue in the downlink of massive \ac{mu} \ac{mimo} \ac{ofdm}, where potentially up to several hundreds of PAs in the \ac{bs} require linearization. In this paper, we propose a \acf{cnn}-based \ac{dpd} in the frequency domain, taking place before the precoding, where the dimensionality of the signal space depends on the number of users, instead of the number of \ac{bs} antennas. Simulation results on \ac{gmp}-based \acp{pa} show that the proposed \ac{cnn}-based DPD can lead to very large complexity savings as the number of \ac{bs} antenna increases at the expense of a small increase in power to achieve the same \ac{ser}.
\end{abstract}

%\begin{IEEEkeywords}
%power amplifiers, digital predistortion, reinforcement learning.
%\end{IEEEkeywords}
%
%%%%%%%%%%%%%%%%%%%%%%%%%%%%%%%%%%%%%%%%%%%%%%%%%%%%%%%%%%%%%%%%%%%%%%%%%%%%
\section{Introduction}
Massive \acf{mu}-\acf{mimo} is one of the prominent technologies in \acf{5g} and beyond. With proper precoding techniques, up to several hundreds of antennas at the \acf{bs} increases the capacity when serving many tens of \acfp{ue}~\cite{larsson2014massive}. However, such large number of antennas poses a difficult linearization problem for the nonlinear \acfp{pa}. To meet the linearization performance of each \ac{pa}, it is common to use \acf{dpd} for each PA. However, the computational complexity increases linearly with the number of \acp{pa} and  becomes unacceptably large  for massive MU-MIMO~\cite{liu2019linearization,wang2019digital}.

To tackle this complexity problem, recent works have shifted toward reducing the number of \acp{dpd}~\cite{liu2019linearization, wang2019digital,liu2016single,yan2017digital}, so that each \ac{dpd} linearizes more than one single \ac{pa}. These methods unavoidably degrade the linearization performance as each PA has a different behavior due to variations in component characteristics. Another research direction goes toward \ac{fd} \ac{dpd}~\cite{brihuega2021frequency, tarver2021virtual}, where \ac{dpd} is implemented before the \ac{ofdm} \ac{idft}. The complexity of the DPD is reduced as the DPD sampling rate becomes equal to the symbol rate, instead of using oversampling as in conventional \ac{dpd} methods. While~\cite{brihuega2021frequency}  considers  hybrid massive MIMO,~\cite{tarver2021virtual} considers a fully digital \ac{mu}-\ac{mimo} system with a \acf{nn}-based DPD, where the \ac{dpd} is implemented prior to the precoder. Thus, the  complexity of the DPD is reduced as the dimensionality of the signal increases with the number of \acp{ue} instead of the number of \ac{bs} antennas. However, this method requires several \acfp{idft} and \acp{dft} as the \ac{nn}-based DPD still operates in the \ac{td}. More problematically, the complexity cost of the method in~\cite{tarver2021virtual} still increases with the number of \ac{bs} antennas due to the additional precoding cost of the guard-band subcarriers. To address this complexity problem, an \ac{fd} \ac{dpd} model that only processes data subcarriers without \acp{idft} would be a possible solution. To the best of our knowledge, such a model has not yet been proposed.
  
In this work, we consider a massive \ac{mu}-\ac{mimo}-\ac{ofdm} system. We propose a   \acf{cnn}-based  \ac{fd} \ac{dpd} model  that takes place before the \ac{ofdm} \ac{idft} and the precoder. We refer to this model as FD-CNN. Compared with conventional \ac{td} per-antenna \ac{dpd}, the complexity cost of FD-CNN increases with the symbol rate and the number of \acp{ue} instead of with the much higher sampling rate and the number of \ac{bs} antennas. Furthermore, to avoid high complexity cost caused by the large number of data subcarriers at the DPD input, we consider \ac{2d} convolutional layers that can efficiently extract information from data subcarriers. We focus on the in-band metric \ac{ser} as the out-of-band metric \ac{acpr} requirements have been considerably relaxed in millimeter wave~\cite{Brihuega_ACPR_2020}.  Simulation results on different \ac{gmp}-based PA models at each antenna and a frequency-selective Rayleigh fading channel show that the proposed \ac{fd}-CNN DPD can provide very large complexity savings when the number of \ac{bs} antennas grows at the expense of a small increase in power to achieve the same \ac{ser} performance compared with  state-of-the-art TD and FD DPDs.

\textit{Notation:} Lowercase and uppercase boldface letters denote column vectors and matrices such as $\boldsymbol{x}$ and $\boldsymbol{X}$; $\boldsymbol{x}^{\mathsf{H}}$ denotes the Hermitian transpose of $\boldsymbol{x}$;  $\mathbb{R}$ and $\mathbb{C}$ denote real and complex numbers, respectively; $x_n$ or $[\boldsymbol{x}]_n$ denote the $n$-th element of $\boldsymbol{x}$, and $\boldsymbol{x}_{n:n+k}$ denotes a vector consisting of the $n$-th to $(n+k)$-th elements of $\boldsymbol{x}$; the $B\times U$ all-zeros matrix and the $U\times U$ identity matrix are denoted by $\boldsymbol{0}_{B\times U}$ and $\boldsymbol{I}_{U}$, respectively; $\mathbb{E}_{\boldsymbol{}}\{x\}$  denotes the expectation of $x$.

%%%%%%%%%%%%%%%%%%%%%%%%%%%%%%%%%%%%%%%%%%%%%%%%%%%%%%%%%%%%%%%%%%%%%%%%%%
\section{System model}
\begin{figure*}[t]
    \centering
    	\vspace*{-0.5 \baselineskip}
    \includegraphics[width=0.8\linewidth]{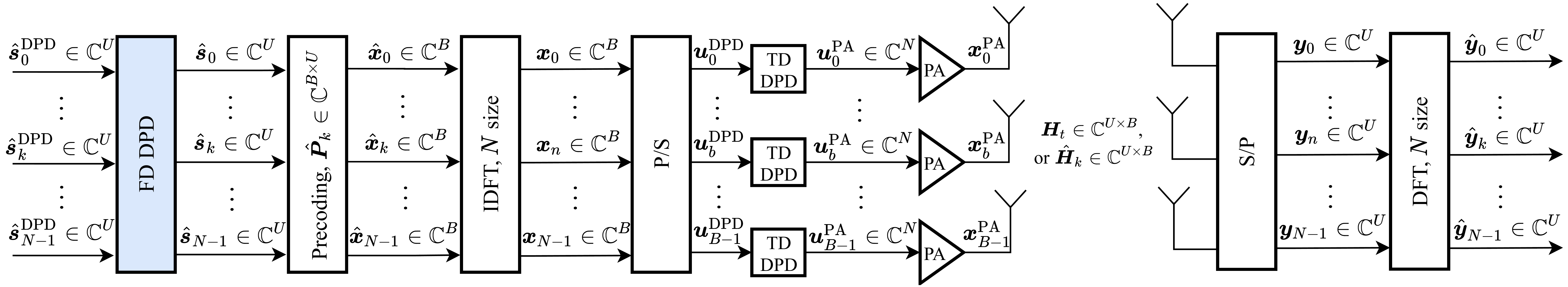}
    \caption{System model of a massive MU-MIMO-OFDM downlink with nonliner PAs in each RF chain of the BS. The conventional TD per-antenna DPDs take place before each PA while the proposed FD DPD operates before the precoder.}
    	\vspace*{-0.5 \baselineskip}
    \label{fig:sys_model_MUMIMO}
\end{figure*}

\subsection{System Model}
We consider a massive MU-MIMO-OFDM downlink system model as shown in Fig.~\ref{fig:sys_model_MUMIMO}. The \ac{bs} is equipped with $B$ antennas and transmits messages to $U \ll B$ single-antenna \acp{ue}. Each OFDM symbol consists of $N$ subcarriers with $N_{\text{d}}$ data subcarriers and $N_{\text{g}}=N-N_{\text{d}}$ guard subcarriers. The subcarrier spacing is denoted by $\Delta f$. Accordingly, the sampling and symbol rate are defined as $f_{\text{s}}=N\Delta f$ and $f_{\text{d}}=
N_{\text{d}}\Delta f$, respectively. The oversampling rate is $R= f_{\text{s}}/f_{\text{d}} = N/N_{\text{d}}$. 

Specifically, let $\hat{\boldsymbol{s}}_k \in \mathbb{C}^{U}$ denote the symbol vector for $U$ \acp{ue} at subcarrier $k$ in the \ac{fd}, where all symbols are generated independently from an  $M$-QAM constellation. A precoder  maps $\hat{\boldsymbol{s}}_k$ to a $B$-dimensional  antenna array  based on the available \ac{csi}. We consider linear precoders due to their low complexity and good performance~\cite{wiesel2008zero}. Given the data vector $\hat{\boldsymbol{s}}_k$ at subcarrier $k$, the \ac{fd} output of a linear precoder, $\hat{\boldsymbol{x}}_k \in \mathbb{C}^{B}$, is obtained by~\cite{jacobsson2019linear}
\begin{align}
    \hat{\boldsymbol{x}}_k = \hat{\boldsymbol{P}}_k \hat{\boldsymbol{s}}_k\,,  
    \label{eq:precod_FD}
\end{align}
where $ \hat{\boldsymbol{P}}_k \in \mathbb{C}^{B\times U}$ denotes the \ac{fd} precoding matrix for subcarrier $k$. We consider the precoding matrices $\hat{\boldsymbol{P}}_k = \boldsymbol{0}_{B\times U}$ for guard subcarriers. The precoded vectors $\hat{\boldsymbol{x}}_k $ are transformed to \ac{td} signals by $N$-size IDFTs. The \ac{td} signal vector at time sample $n$, $\boldsymbol{x}_n \in \mathbb{C}^{B}$, is given by
\begin{align}
    \boldsymbol{x}_n=\frac{1}{\sqrt{N}}\sum_{k=0}^{N-1} \hat{\boldsymbol{x}}_k \exp\left(jk\frac{2\pi}{N}n\right)\,. 
    \label{eq:IDFT_precod}
\end{align}

Without the impact of PA imperfections, the received signal at the $U$ UEs in the \ac{td} at time sample $n$, $\boldsymbol{y}_n \in \mathbb{C}^{U}$, is given by~\cite{jacobsson2019linear}
\begin{align}
    \boldsymbol{y}_n = \sum_{t=0}^{T-1}\boldsymbol{H}_t\boldsymbol{x}_{n-t} + \boldsymbol{w}_n\,,
    \label{eq:channel_TD}
\end{align}
where $\boldsymbol{H}_t \in \mathbb{C}^{U\times B}$ denotes the \ac{td} channel matrix at the $t$-th tap with a total of $T$ taps. The elements of the channel matrix are independently generated from a complex Gaussian distribution with zero mean and variance $T^{-1}$~\cite{jacobsson2019linear}. The channel is assumed to be block-constant with a coherence time of one OFDM symbol. In addition, 
 $\boldsymbol{w}_{n} \sim \mathcal{C} \mathcal{N}\left(\boldsymbol{0}_{U \times 1}, N_0 \boldsymbol{I}_{U}\right)$ denotes the \ac{awgn} vector at time sample $n$ with the noise variance $N_0$. Note that this yields a spatially white frequency-selective Rayleigh-fading channel with uniform power-delay profile~\cite{jacobsson2019linear}. 

Equivalently,  \eqref{eq:channel_TD} can be written in the \ac{fd} as
\begin{align}
    \hat{\boldsymbol{y}}_k = \hat{\boldsymbol{H}}_k \hat{\boldsymbol{x}}_k + \hat{\boldsymbol{w}}_k,
    \label{eq:channel_FD}
\end{align}
where $\hat{\boldsymbol{H}}_k \in \mathbb{C}^{U\times B} = \sum_{t=0}^{T-1} \boldsymbol{H}_t \exp(-jk\frac{2\pi}{N}t)$ is the \ac{fd} channel matrix at subcarrier $k$, 
$\hat{\boldsymbol{y}}_k \in \mathbb{C}^{U} = \sum_{n=0}^{N-1} \boldsymbol{y}_n \exp(-jk\frac{2\pi}{N}n)$ is the \ac{fd} received signal at subcarrier $k$, and $\hat{\boldsymbol{w}}_k \in \mathbb{C}^{U} = \sum_{n=0}^{N-1} \boldsymbol{w}_n \exp(-jk\frac{2\pi}{N}n)$ is the \ac{fd} AWGN noise at subcarrier $k$. Substituting~\eqref{eq:precod_FD} into~\eqref{eq:channel_FD}, $\hat{\boldsymbol{y}}_k$ with precoder $\hat{\boldsymbol{P}}_k$ can be expressed as
\begin{align}
    \hat{\boldsymbol{y}}_k = \hat{\boldsymbol{H}}_k \hat{\boldsymbol{P}}_k \hat{\boldsymbol{s}}_k + \hat{\boldsymbol{w}}_k\,.
    \label{eq:channel_FD_with_precoding}
\end{align}

To minimize the MU interference, the pseudo-inverse of the channel matrix is used as the precoding matrix, which is known as the \ac{zf} precoding. The precoding matrix $\hat{\boldsymbol{P}}_k$ in~\eqref{eq:channel_FD_with_precoding} can be expressed as~\cite{wiesel2008zero}
\begin{align}
    \hat{\boldsymbol{P}}_k = \alpha \hat{\boldsymbol{H}}_k^{\mathsf{H}}(\hat{\boldsymbol{H}}_k \hat{\boldsymbol{H}}_k^{\mathsf{H}})^{-1},
    \label{eq:Zero_forcing}
\end{align}
where $\alpha$ denotes the normalization factor to ensure the power constraint $\mathbb{E}\left\{\|\hat{\boldsymbol{x}}_{k}\|^{2}\right\}=P_{T}$ is met. The expectation is over the symbols of all UEs, and $P_T$ denotes the average transmit power.
With an imperfect knowledge of the \ac{csi}, the BS has an estimated \ac{td} channel matrix $\boldsymbol{H}_t^{\text{est}} = \sqrt{1-\eta}\boldsymbol{H}_t + \sqrt{\eta}\boldsymbol{H}_t^{\text{err}}$, where $\eta \in [0,1]$ and $\boldsymbol{H}_t^{\text{err}} \sim \mathcal{C} \mathcal{N}\left(\mathbf{0}_{U \times B}, \boldsymbol{I}_{U\times B}\right)$~\cite{jacobsson2019linear}. Thus, the \ac{fd} channel matrix in~\eqref{eq:Zero_forcing} is replaced by a channel estimation $\hat{\boldsymbol{H}}_k^{\text{est}}= \sum_{t=0}^{T-1} \boldsymbol{H}_{t}^{\text{est}} \exp \left(-j k \frac{2 \pi}{N} t\right)$.

\subsection{PA nonlinearity}
After the IDFT, the \ac{td} OFDM symbols are mapped to $B$ \ac{rf} chains and sent to each \ac{td} DPD with input at the $b$-th RF chain, $\boldsymbol{u}_b^{\text{DPD}} \in \mathbb{C}^{N} = [[\boldsymbol{x}_0]_{b},[\boldsymbol{x}_1]_{b},...,[\boldsymbol{x}_{N-1}]_{b}]^{\mathsf{T}}$. The DPD outputs are then converted to analog domain by the \acp{dac}. For simplicity, we assume ideal \acp{dac} with infinite-resolution.    The PA input signal at the $b$-th RF chain is $\boldsymbol{u}_b^{\text{PA}} \in \mathbb{C}^{N}$.\footnote{\vspace*{-0\baselineskip} In Section II-C, we describe the input-output relation of TD DPD and how the PA input $\boldsymbol{u}_b^{\text{PA}}$ is obtained. \vspace*{-0\baselineskip}} The PA associated with the $b$-th BS antenna is represented by the nonlinear function $f^{\mathrm{PA}}_b: \mathbb{C}^{L_1+1} \rightarrow \mathbb{C}$ with memory length $L_1$, input  $[\boldsymbol{u}_b^{\text{PA}}]_{n: n-L_{1}}=\left[[\boldsymbol{u}_b^{\text{PA}}]_n, \ldots, [\boldsymbol{u}_b^{\text{PA}}]_{n-L_1}\right]^{\mathsf{T}} \in \mathbb{C}^{L_{1}+1}$, and output $x^{\text{PA}}_{b,n}$. The input-output relation of the $b$-th PA at time sample $n$ can be expressed as
\begin{align}
    x^{\text{PA}}_{b,n} = f^{\mathrm{PA}}_b([\boldsymbol{u}_b^{\text{PA}}]_{n: n-L_{1}}).
    \label{eq:PA_b}
\end{align}
Assuming an ideal PA with a linear behavior and with no need of DPD, \eqref{eq:PA_b} reduces to $x^{\text{PA}}_{b,n} = [\boldsymbol{u}_b^{\text{PA}}]_n = [\boldsymbol{u}_b^{\text{DPD}}]_n = [\boldsymbol{x}_n]_b$, where $[\boldsymbol{x}_n]_b$ is the $b$-th element of $\boldsymbol{x}_n$.  With the PA gain, the average PA output power is defined as $P_{\text{PA}} = \mathbb{E}\{|x^{\text{PA}}_{b,n}| \}$, where the expectation is over all time steps and BS antennas. In this case, no \ac{ici} is introduced, and MU interference is eliminated by the \ac{zf} precoding, given perfect CSI. However, the PA nonlinearity creates distortion which in FD breaks the orthogonality of the subcarriers, and in spatial domain breaks the nulls created by the \ac{zf} precoding. Thus, the nonlinear PAs cause both \ac{ici} and MU interference.

\subsection{Time-domain DPD}
DPD is applied to compensate for the nonlinear behavior of the PA. Conventionally, DPD is implemented in  \ac{td} before the PA and makes use of oversampling in order to cancel out adjacent channel power created by the PA nonlinearity. We represent the TD DPD associated with the $b$-th BS PA  by a parametric model $f_{\boldsymbol{\theta}^{\text{DPD}}_b}: \mathbb{C}^{L_2+1} \rightarrow \mathbb{C}$ with parameters $\boldsymbol{\theta}_{b}^{\text{DPD}}$ and input memory length $L_2$. Then, the mapped analog \ac{td} signal $\boldsymbol{u}_b^{\text{DPD}}$ is sent to the corresponding DPD as 
\begin{align}
    u^{\text{PA}}_{b,n} = f_{\boldsymbol{\theta}^{\text{DPD}}_b}([\boldsymbol{u}_b^{\text{DPD}}]_{n: n-L_{2}}),
    \label{eq:TD_DPD_b}
\end{align}
where the DPD output $u^{\text{PA}}_{b,n}$ corresponds to the PA input in~\eqref{eq:PA_b}. In the case of massive MU-MIMO, one can use one DPD for  each PA. However, the total computational complexity cost of the DPD increases linearly with the number of PAs and antennas, which makes this approach highly impractical when several hundreds of PAs are deployed~\cite{liu2019linearization}.

\section{Complexity Analysis of DPD Benchmarks in Massive MU-MIMO}
\subsection{Measure of DPD Complexity}
When working with DPD algorithm design, there are two complexity figures to consider, the \textit{running complexity} and \textit{traning complexity}. The \textit{running complexity} is the computational complexity needed to continuously run the DPD algorithm, which makes it the major overall contributor to complexity. Unlike the \textit{training complexity}, which is an offline cost, the running complexity directly relates to the number of operations required for the inference of a certain number of samples~\cite{complexityGMP}. 

There are several ways to measure the running complexity of DPD such as the Bachmann-Landau measure $\mathcal{O}(\cdot)$, running time, and number of parameters. These measures are either approximations or depend  on the specific type of platform and thus are implementation-dependent. In contrast, the number of \acp{flop} can accurately measure every addition, subtraction, and multiplication operation for both Volterra series-based models~\cite{moon2011enhanced} and NNs~\cite{liu2004dynamic, wu2021low} for DPD. We therefore utilize the number of \acp{flop} to measure the running complexity of DPD, as this is the figure of merit which brings us closest to a fair comparison without going into implementation details. In this paper, we follow the same FLOP calculation as in~\cite[Table I]{complexityGMP}. In massive MU-MIMO, for a fair complexity comparison for different DPD schemes in the \ac{td} and \ac{fd}, we calculate the total number of FLOPs required per \ac{ue} for one OFDM symbol.

\subsection{Time-domain Per-antenna GMP-based DPD}
While there exist a large variety of published \ac{td} DPD models, we choose the commonly used \ac{gmp}~\cite{GMP_2006} for comparison because it has been shown to outperform many other models in terms of linearization performance versus complexity~\cite{complexityGMP}. For simplicity, we refer to this \ac{td} per-antenna GMP method as TD-GMP.

To linearize every PA in massive MU-MIMO-OFDM, each RF chain deploys a GMP-based DPD with the same memory length $L_{3}$, nonlinear order $K$, and cross-term length $G$. Given the \ac{td} signal, $\boldsymbol{u}_b$, at the $b$-th RF chain as the input, the output of the \ac{gmp}-based DPD associated with the $b$-th RF chain at time sample $n$ can be expressed as
\begin{equation}
    \begin{aligned}
u^{\text{PA}}_{b,n}=& \sum_{k=0}^{K-1} \sum_{l=0}^{L_3} a_{k,l} [\boldsymbol{u}_b^{\text{DPD}}]_{n-l}\big| [\boldsymbol{u}_b^{\text{DPD}}]_{n-l}\big|^{k} \\
&+\sum_{k=1}^{K-1} \sum_{l=0}^{L_3} \sum_{g=1}^{G}\Big(b_{k,l,g} [\boldsymbol{u}_b^{\text{DPD}}]_{n-l}\big|[\boldsymbol{u}_b^{\text{DPD}}]_{n-l-g}\big|^{k}\\
&+c_{k,l,g} [\boldsymbol{u}_b^{\text{DPD}}]_{n-l}\left|[\boldsymbol{u}_b^{\text{DPD}}]_{n-l+g}\right|^{k}\Big)\,,
\end{aligned}
\label{eq:GMP_b}
\end{equation}
where $a_{k,l}$, $b_{k,l,g}$, and $c_{k,m,g}$ are complex-valued coefficients, which are identified using conventional \ac{mse} estimation. 

The number of \acp{flop} required for the GMP with each input sample is computed as~\cite{complexityGMP}
\begin{align}
    %\begin{aligned}
C^{\text{TD-GMP}}_{\text{Samp}}&= 8\big((L_3+1)(K+2 K G)-\frac{G(G+1)}{2}(K-1)\big) \notag \\
&+10+2 K+2(K-1) G+2 K \min (G, L_3) \label{eq:GMP_FLOPs_per_sample}
%\\& \approx \mathcal{O}\left(L_3 K G \right)\,. \notag 
%\end{aligned}
\end{align}
Now consider that we use GMP-based DPD for each antenna in a massive MU-MIMO antenna system. The number of FLOPs required per \ac{ue} for one OFDM symbol is calculated as
\begin{align}
   C^{\text{TD-GMP}}_{\text{}} = C^{\text{TD-GMP}}_{\text{Samp}} NB/U\,.
   \label{eq:TD-GMP-Complexity}
\end{align}
 We note that the number of FLOPs for the \ac{td} GMP-based DPD grows linearly with the number of \ac{bs} antennas because every PA requires one DPD. This causes a complexity problem in massive MU-MIMO with hundreds of antennas.
 
 \begin{figure*}[t!]
    \centering
    	\vspace*{0 \baselineskip}
    \includegraphics[width=0.55\linewidth]{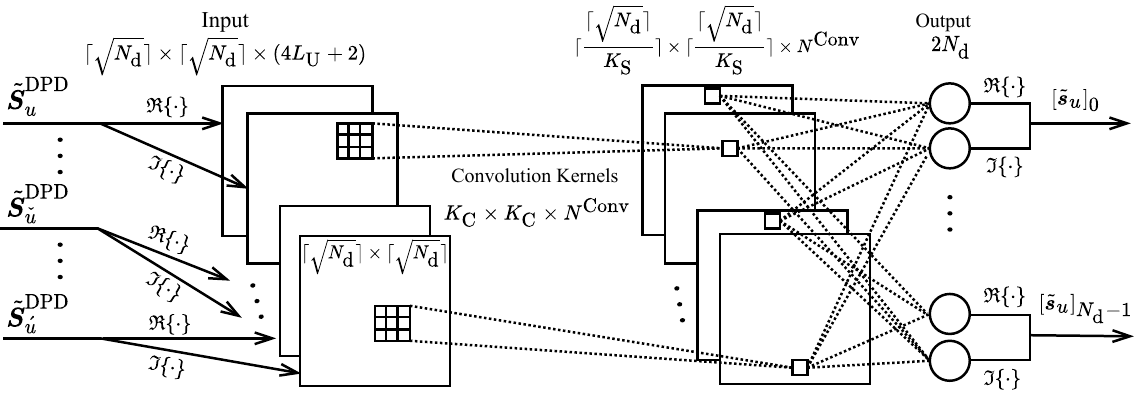}
    \caption{Structure of the proposed FD-CNN. For the $u$-th UE, the input of FD-CNN is formed by a selection of $L_{\text{U}}$ UE symbol vectors, $\{\tilde{\boldsymbol{s}}_u^{\text{DPD}}, \tilde{\boldsymbol{s}}_{\check{u}}^{\text{DPD}}, ..., \tilde{\boldsymbol{s}}_{\acute{u}}^{\text{DPD}}\}$. The output is the predistorted symbol vector of $u$-th  UE, $\tilde{\boldsymbol{s}}_{u}$.}
    \label{fig:FD-CNN-diagram}
    	\vspace*{-0.5 \baselineskip}
\end{figure*}

\subsection{\ac{fd} Neural Network-based DPD}
To address the complexity problem of \ac{td} DPD in massive MU-MIMO, the authors in~\cite{tarver2021virtual} proposed a \ac{nn}-based DPD, which operates in the \ac{fd} prior to the precoder. In this paper,  we refer to this \ac{fd} NN-based DPD model as FD-NN. To form the input of the FD-NN, each \ac{ue} stream is first converted to the \ac{td}. Specifically, given the symbol vector for the $u$-th \ac{ue} in the \ac{fd}, $\tilde{\boldsymbol{s}}_u^{\text{DPD}} \in \mathbb{C}^{N} = [[\hat{\boldsymbol{s}}_0^{\text{DPD}}]_{u},[\hat{\boldsymbol{s}}_{1}^{\text{DPD}}]_{u},...,[\hat{\boldsymbol{s}}_{N-1}^{\text{DPD}}]_{u}]^{\mathsf{T}}$, the \ac{td} symbol vector for the $u$-th \ac{ue}, $\boldsymbol{s}_u^{\text{DPD}} \in \mathbb{C}^{N}$, is given by an IDFT
\begin{align}
    \boldsymbol{s}_u^{\text{DPD}} = \frac{1}{\sqrt{N}}\sum_{k=0}^{N-1} [\tilde{\boldsymbol{s}}_u^{\text{DPD}}]_{k} \exp(jk\frac{2\pi}{N}n) = \boldsymbol{F}_N^{\mathsf{H}}\tilde{\boldsymbol{s}}_u^{\text{DPD}},
\end{align}
where $\boldsymbol{F}_N$ denotes the unitary $N\times N$ DFT matrix. Then, using tapped delay lines with memory length $L_4$, the input signal of the first layer consists of each \ac{td} UE symbol vector, $[\boldsymbol{s}_u^{\text{DPD}}]_{n:n-L_4} = [[\boldsymbol{s}_u^{\text{DPD}}]_{n},[\boldsymbol{s}_u^{\text{DPD}}]_{n-1},...,[\boldsymbol{s}_u^{\text{DPD}}]_{n-L_4}]^{\mathsf{T}} \in \mathbb{C}^{L_4+1}$. In total, the input signal of the first layer FD-NN is given by
\begin{align}
    \boldsymbol{s}_{\text{In}}^{\text{FD-NN}} = [[\boldsymbol{s}_0^{\text{DPD}}]_{n:n-L_4}^{\mathsf{T}}, [\boldsymbol{s}_1^{\text{DPD}}]_{n:n-L_4}^{\mathsf{T}},...,[\boldsymbol{s}_{U-1}^{\text{DPD}}]_{n:n-L_4}^{\mathsf{T}}]^{\mathsf{T}},
\end{align}
which is then decomposed into real and imaginary parts connecting with their own neurons. Thus, this yields $2U(L_4+1)$  neurons for the first layer.
All layers in the FD-NN are fully connected. There are $K^{\text{NN}}$ hidden layers, each with $D$ neurons and a nonlinear activation function ReLU. The number of neurons at the output layer is $2U$. The predistorted \ac{td} signal for each UE is then converted back to the \ac{fd} by \acp{dft} before being sent to the precoder. 

In this method, the additional \acp{idft} and \acp{dft} introduce an extra complexity cost. More importantly, the guard subcarriers of the predistorted signal are not empty anymore due to the \acp{dft}, which leads to extra complexity cost due to the precoding. 
For each UE, the number of \acp{flop} required for the FD-NN per OFDM symbol can be computed as
\begin{align}
%\begin{aligned}
        C_{\text{}}^{\text{FD-NN}} & = \frac{1}{U} \Big( \underbrace{N\big(4U (L_3+1) D  + 2(K^{\text{NN}}-1) D^2 + 4DU\big)}_{\text{NN}} \notag \\ &\quad +  \underbrace{10UN\log_2N}_{\text{(I)DFTs}} + \underbrace{(6+2)N_gUB}_{\text{Extra precoding}} \Big), \label{eq:FD-NN-Complexity}
        %\\
        %& \approx \mathcal{O}( N L_3 D + K^{\text{NN}}D^2/U + N\log_2N + N_gB) \notag
%\end{aligned}
\end{align}
where  the $6$ and $2$ in the extra precoding part are for complex-number multiplication and addition, respectively.  Considering the fast Fourier transform, the complexity of a $N$-size (I)DFT is approximated by $5N\log_2N$ FLOPs with $\frac{N}{2}\log_2N$ complex-number multiplications and $N\log_2N$ complex-number additions, and there are $U$ IDFTs and $U$ DFTs. We note that due to extra precoding cost, the number of FLOPs for FD-NN still increases with the number of BS antennas, which dominates the complexity when hundreds of antennas are deployed in massive MU-MIMO.

\section{Proposed \ac{fd} Convolutional Neural Network} 

\subsection{Structure of the FD-CNN}
The structure of the proposed FD-CNN is shown in Fig.~\ref{fig:FD-CNN-diagram}. FD-CNN works in the \ac{fd} with the input of UE symbol vectors. To efficiently extract appropriate information from the \ac{fd} UE symbol vectors without using IDFTs as~\cite{brihuega2021frequency},  we propose to use \ac{2d} convolutional layers as the first  layer, which save complexity  compared with a fully connected layer. To facilitate the 2D convolution, the \ac{fd} UE symbol vector $\tilde{\boldsymbol{s}}_u^{\text{DPD}} \in \mathbb{C}^{N}$ is converted to a matrix $\tilde{\boldsymbol{S}}_u^{\text{DPD}} \in \mathbb{C}^{\lceil\sqrt{N_{\text{d}}}\rceil \times \lceil\sqrt{N_{\text{d}}} \rceil}$,\footnote{\vspace*{0\baselineskip}The vector-to-matrix conversion follows a contiguous order so that the convolution kernels can efficiently extract information from adjacent subcarriers. \vspace*{-0\baselineskip}} where only $N_{\text{d}}$ data subcarriers are involved to save complexity. Here $\lceil\cdot\rceil$ denotes the ceiling function and zero-padding is used if $\lceil\sqrt{N_{\text{d}}} \rceil> \sqrt{N_{\text{d}}}$. To offer more complexity saving, for the $u$-th UE, an arbitrary selection of $L_{\text{U}}$ UE symbol vectors from all $U$ UEs, $\{\tilde{\boldsymbol{s}}_u^{\text{DPD}}, \tilde{\boldsymbol{s}}_{\check{u}}^{\text{DPD}}, ..., \tilde{\boldsymbol{s}}_{\acute{u}}^{\text{DPD}}\}$, including $\tilde{\boldsymbol{s}}_u^{\text{DPD}}$, are converted to matrices.   In total, these matrices form the input of the FD-CNN as a tensor, $\tilde{\mathcal{S}}_{u} \in \mathbb{C}^{\lceil\sqrt{N_{\text{d}}}\rceil \times \lceil\sqrt{N_{\text{d}}} \rceil\times L_{\text{U}}}$,
\begin{align}
	\tilde{\mathcal{S}}_{u}^{\text{DPD}} = [\tilde{\boldsymbol{S}}_{u}^{\text{DPD}},...,\tilde{\boldsymbol{S}}_{\check{u}}^{\text{DPD}},...,\tilde{\boldsymbol{S}}_{\acute{u}}^{\text{DPD}} ].
	\label{eq:FD-CNN-input}
\end{align}
 Then, each complex-valued UE symbol matrix in $\tilde{\mathcal{S}}_{u}^{\text{DPD}}$ is decomposed into real and imaginary matrices with the same dimension as $\tilde{\boldsymbol{S}}_{u}^{\text{DPD}}$. In total, there are $2L_{\text{U}}$ real-valued  UE symbol matrices, which are convoluted by $N^{\text{Conv}}$ convolutional kernels with the same kernel size $K_{\text{C}}$ and stride length $K_{\text{S}}$. This yields an output tensor, $\tilde{\mathcal{S}}_{u}^{\text{Conv}} \in \mathbb{R}^{\lceil\frac{\lceil\sqrt{N_{\text{d}}}\rceil}{K_{\text{S}}}\rceil \times \lceil\frac{\lceil\sqrt{N_{\text{d}}}\rceil}{K_{\text{S}}}\rceil \times N^{\text{Conv}}}$, where zero-padding is also considered if $\lceil\frac{\lceil\sqrt{N_{\text{d}}}\rceil}{K_{\text{S}}}\rceil > \frac{\lceil\sqrt{N_{\text{d}}}\rceil}{K_{\text{S}}}$.
 
 $\tilde{\mathcal{S}}_{u}^{\text{Conv}}$ is passed through an element-wise activation function $\sigma(\cdot)$ and then flatten into a column vector, $\tilde{\boldsymbol{s}}_{u}^{\text{Flat}}\in \mathbb{R}^{\lceil\frac{\lceil\sqrt{N_{\text{d}}}\rceil}{K_{\text{S}}}\rceil \lceil\frac{\lceil\sqrt{N_{\text{d}}}\rceil}{K_{\text{S}}}\rceil N^{\text{Conv}}}$. After that  $\tilde{\boldsymbol{s}}_{u}^{\text{Flat}}$ is fully connected with the linear output layer with $2N_{\text{d}}$ neurons, which correspond to the real and imaginary parts of the  predistorted symbol vector at the $u$-th UE, $\tilde{\boldsymbol{s}}_u \in \mathbb{C}^{N_{\text{d}}}$.
\subsection{Complexity Analysis}
We split the complexity of the proposed FD-CNN into the convolution layer part and the fully-connected layer part. For the complexity of the convolution layer part, each $K_{\text{C}}$-size 2D convolution requires $(2K_{\text{C}}^2 -1)$ \acp{flop} consisting of $K_{\text{C}}^2$ real-valued multiplications and $(K_{\text{C}}^2-1)$ real-valued additions. This 2D convolution operates on $2L_{\text{U}}$ real-valued UE symbol matrices with stride length $K_{\text{S}}$, and in total there are $N^{\text{Conv}}$ convolutional kernels. Thus, the number of \acp{flop} for the convolution layer part can be expressed as
\begin{align}
	C_{\text{Conv}}^{\text{FD-CNN}} = 2N^{\text{Conv}}L_{\text{U}} (2K_{\text{C}}^2 -1) \left(\frac{\lceil\sqrt{N_{\text{d}}}\rceil}{K_{\text{S}}}\right)^2. 
\end{align}

The number of FLOPs for the fully-connected layer part can be expressed as
\begin{align}
	C_{\text{FC}}^{\text{FD-CNN}} =  8N_{\text{d}}L_{\text{U}} \left(\frac{\lceil\sqrt{N_{\text{d}}}\rceil}{K_{\text{S}}}\right)^2.
\end{align}

In total, the number of \acp{flop} required for the FD-CNN with one OFDM symbol per UE is calculated as
\begin{align}
C^{\text{FD-CNN}} & = C_{\text{Conv}}^{\text{FD-CNN}} + C_{\text{FC}}^{\text{FD-CNN}}
	\label{eq:FD-CNN-Complexity}
	%\\
	%& \approx \mathcal{O}(N^{\text{Conv}}L_{\text{U}}K^2_{\text{C}}N_{\text{d}} /K^2_{\text{S}} +N_{\text{d}}L_{\text{U}}N_{\text{d}}/K^2_{\text{S}})
\end{align}
We note that $C^{\text{FD-CNN}}$ is sensitive to the number of data subcarriers $N_{\text{d}}$ because $C_{\text{FC}}^{\text{FD-CNN}}$ is in the magnitude of $N_{\text{d}}^2$. Compared with the complexity for TD-GMP and FD-NN in~\eqref{eq:TD-GMP-Complexity} and~\eqref{eq:FD-NN-Complexity}, we note that the complexity of FD-CNN does not depend on the number of BS antennas, $B$, which naturally saves complexity  when $B$ is large in massive MU-MIMO. 

\section{Numerical Results}

\subsection{Simulation Setup}
\subsubsection{Parameters}
We consider a set of simulated PAs using the GMP model in~\eqref{eq:GMP_b} with nonlinear order $K=7$, memory length $M=3$, and cross-term length $G=1$. These parameters are estimated using real measurements from the RF WebLab using a $100$ MHz OFDM signal~\cite{landin2015weblab}. Using the estimated PA parameters as the mean of a Gaussian distribution, the PA parameters of each antenna are drawn with variance $0.01$. For instance, given the original GMP-based PA coefficient $\alpha_{k,l}$, the coefficient at $b$-th antenna $\alpha_{k,l}^{b} \sim \mathcal{N}(\alpha_{k,l}, 0.01\alpha_{k,l})$. Note the GMP-based PA coefficients are fixed once being generated.  The saturation point and measurement noise standard deviation of each PA are $24.02$ V $(\approx 37.6$ dBm with a $50$ $\Omega$ load impedance) and $0.053$ V, respectively. 

We select a $50$ MHz OFDM setup in the operating band n257 from the 3GPP 5G standardization~\cite[Table 5.3.2-1 and 5.3.5-1]{3gpp.36.331} with subcarrier spacing $\Delta f=120$ kHz, OFDM symbol length $N=4096$, number of data subcarriers $N_{\text{d}}=384$, and $M=256$ QAM modulation. Each channel realization has $T=10$ taps. The AWGN noise variance is $N_0=1$ Watts/Hz. We assume the BS has imperfect CSI with $\eta=0.001$, which affects the ZF precoding.

For the benchmark of TD-GMP, we consider the same GMP parameter setup for each DPD as the PA model, namely $K=7$, $L_3=3$, and $G=1$. For the benchmark of FD-NN, we consider the same memory length $L_4=3$.  Other parameters are set the same as in~\cite{tarver2021virtual} with $N_{\text{FD-NN}}=1$ hidden layer and $D=15$ neurons. For the proposed FD-CNN, we set $\left\lceil\sqrt{N_{\mathrm{d}}}\right\rceil=20$, $L_{\text{U}}=1$, $K_{\text{C}}=3$, and $K_{\text{S}}=1$, and $N^{\text{Conv}}=2$. The activation function $\sigma(\cdot)$ in the convolution layer is the ReLU. To preserve the same input and output dimension of the convolutional layer, zero padding is used. 

\subsubsection{DPD Coefficients Identification}
For the TD-GMP, the coefficients of each GMP-based DPD are estimated using the \ac{ila} at the output of each PA~\cite{eun1997new}, where the least squares algorithm is used to minimize the \ac{mse} between the PA input signal and post-distorter output signal. For the FD-NN, since all blocks in the given communication system are simulated and differentiable, we simply utilize supervised learning with a loss function at the receiver side to minimize the \ac{mse} between received symbols $\hat{\boldsymbol{y}}_k$ and transmitted symbols $\hat{\boldsymbol{s}}_k$ for all data subcarriers. This symbol-based criterion for DPD optimization has been shown to achieve similar SER performance in~\cite{wu2021symbol}. Note new channel realization is generated for every training mini-batch. The FD-NN is then trained till convergence using the gradient descent optimizer Adam~\cite{kingma2014adam} with a learning rate $0.001$. 

For the proposed FD-CNN, we choose the same supervised learning, loss function, and optimizer Adam as the training for FD-NN. Each mini-batch consists of $10$ OFDM symbols. In practical applications, it is straightforward to use a  training method similar to the one  in~\cite{tarver2021virtual}. This training method creates the training data of the NN output by converting back PA output signals through MP-based DPDs and the pseudo-inverse of precoders. This requires dedicated feedback paths to collect each PA output at the transmitter. Alternatively, one can use over-the-air method that utilize an observation receiver for data acquisition as in~\cite{hausmair2018modeling,wu2021symbol}.
\subsection{Simulation results}

\subsubsection{Complexity Versus Number of BS Antennas}
\begin{figure}[t]
	\centering
		\vspace*{0.3 \baselineskip}
	% This file was created by tikzplotlib v0.9.8.
\begin{tikzpicture}[font=\scriptsize]
\definecolor{color0}{rgb}{0.12156862745098,0.466666666666667,0.705882352941177}
\definecolor{color1}{rgb}{1,0.498039215686275,0.0549019607843137}
\definecolor{color2}{rgb}{0.172549019607843,0.627450980392157,0.172549019607843}
\definecolor{color3}{rgb}{0.83921568627451,0.152941176470588,0.156862745098039}
\definecolor{color4}{rgb}{0.580392156862745,0.403921568627451,0.741176470588235}
\definecolor{color5}{rgb}{0.549019607843137,0.337254901960784,0.294117647058824}
\definecolor{color6}{rgb}{0.890196078431372,0.466666666666667,0.76078431372549}
\definecolor{color7}{rgb}{0.737254901960784,0.741176470588235,0.133333333333333}

\begin{axis}[
width=7.5cm,
height=5.5cm,
legend cell align={left},
legend style={
  fill opacity=1,
  draw opacity=1,
  text opacity=1,
  at={(0,1)},
  anchor=north west
},
log basis x={2},
log basis y={10},
tick align=outside,
tick pos=left,  
x grid style={white!69.0196078431373!black},
xlabel={Number of antennas, $B$},
xmajorgrids,
xmin=1, xmax=1024,
xmode=log,
xtick={1,4,16,64,256,1024},
xticklabels={1,4,16,64,256,1024},
%minor xtick={0,0.2,...,21},
xtick style={color=black},
y grid style={white!69.0196078431373!black},
ylabel={Number of FLOPs},
ymajorgrids,
yminorgrids,
ymin=214697.188550675,
ymax=4330672033.71046,
ymode=log,
ytick style={color=black}
]
\addplot [semithick, color0, line width=1.0pt, mark=o, mark size=2, mark options={solid,fill opacity=0}]
table {%
1 2695168
2 5390336
4 10780672
8 21561344
16 43122688
32 86245376
64 172490752
128 344981504
256 689963008
512 1379926016
1024 2759852032
};
\addlegendentry{$U=1$, $C^{\text{TD-GMP}}$~\cite{GMP_2006}}
\addplot [semithick, color1, line width=1.0pt, mark=o, mark size=2, mark options={solid,fill opacity=0}]
table {%
1 673792
2 1347584
4 2695168
8 5390336
16 10780672
32 21561344
64 43122688
128 86245376
256 172490752
512 344981504
1024 689963008
};
\addlegendentry{$U=4$, $C^{\text{TD-GMP}}$~\cite{GMP_2006}}
\addplot [semithick, color2, line width=1.0pt, mark=o, mark size=2, mark options={solid,fill opacity=0}]
table {%
1 336896
2 673792
4 1347584
8 2695168
16 5390336
32 10780672
64 21561344
128 43122688
256 86245376
512 172490752
1024 344981504
};
\addlegendentry{$U=8$, $C^{\text{TD-GMP}}$~\cite{GMP_2006}}
\addplot [semithick, color3, mark=square, mark size=2, line width=1.0pt, mark options={solid,fill opacity=0}]
table {%
1 1750016
2 1779712
4 1839104
8 1957888
16 2195456
32 2670592
64 3620864
128 5521408
256 9322496
512 16924672
1024 32129024
};
\addlegendentry{$U\in\{1,4,8\}$, $C^{\text{FD-NN}}$~\cite{tarver2021virtual}}
\addplot [semithick, color4, line width=1.0pt, mark=triangle, mark size=2.5, mark options={solid,fill opacity=0}]
table {%
1 1205760
2 1205760
4 1205760
8 1205760
16 1205760
32 1205760
64 1205760
128 1205760
256 1205760
512 1205760
1024 1205760
};
\addlegendentry{$U\in\{1,4,8\}$, $C^{\text{FD-CNN}}$}
\end{axis}

\end{tikzpicture}
			\vspace*{-0.5 \baselineskip}
	\caption{The number of FLOPs required for each DPD scheme with one UE and one OFDM symbol versus the number of BS antennas, $B$. }
	\label{fig:FLOPs_vs_B}
	\vspace*{-0.5 \baselineskip}
\end{figure}
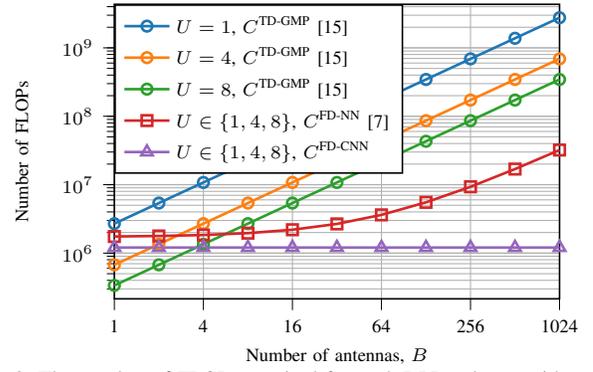
Fig.~\ref{fig:FLOPs_vs_B} shows the number of FLOPs for one OFDM symbol per UE as a function of the number of BS antennas, $B\in\{1,2,4,8,16,32,64,128,256,512,1024\}$, for DPD schemes of TD-GMP, FD-NN, and the proposed FD-CNN. We also plot results for different number of UEs, $U\in \{1,4,8\}$. 

We note that, as $B$ increases, the number of FLOPs for TD-GMP grows linearly as each antenna is associated with one \ac{td} DPD. While the number of FLOPs for FD-NN grows gently for $B<16$, it eventually increases linearly with $B$ as the extra cost of guard-band precoding starts to dominate. However, the number of FLOPs for the proposed FD-CNN is completely invariant to $B$, which would greatly benefit very large antenna systems. Compared with FD-NN and TD-GMP, FD-CNN saves around $2.2\times$ and $8.9\times$ FLOPs for $B=32$ and $U=8$, respectively, and these savings can be further increased up to $26\times$ and $286\times$ when $B$ increases to $1024$.

We also observe that as $U$ increases, the number of FLOPs per UE decreases accordingly for TD-GMP, while this number for FD-NN and FD-CNN is the same.

\subsubsection{SER Versus Average PA Output Power}
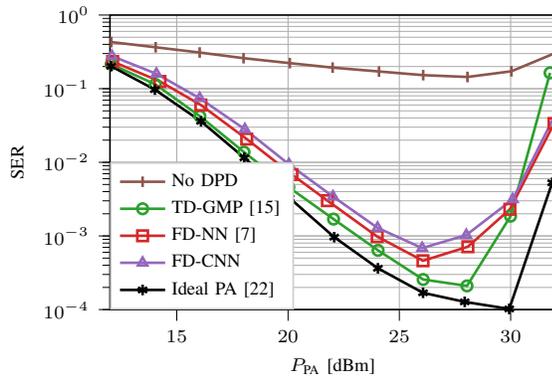
\begin{figure}[t]
	\centering
	% This file was created by tikzplotlib v0.9.8.
\begin{tikzpicture}[font=\scriptsize]

\definecolor{color0}{rgb}{0.12156862745098,0.466666666666667,0.705882352941177}
\definecolor{color1}{rgb}{1,0.498039215686275,0.0549019607843137}
\definecolor{color2}{rgb}{0.172549019607843,0.627450980392157,0.172549019607843}
\definecolor{color3}{rgb}{0.83921568627451,0.152941176470588,0.156862745098039}
\definecolor{color4}{rgb}{0.580392156862745,0.403921568627451,0.741176470588235}
\definecolor{color5}{rgb}{0.549019607843137,0.337254901960784,0.294117647058824}
\definecolor{color6}{rgb}{0.890196078431372,0.466666666666667,0.76078431372549}
\definecolor{color7}{rgb}{0.737254901960784,0.741176470588235,0.133333333333333}
\begin{axis}[
width=7.5cm,
height=5.5cm,
legend cell align={left},
legend style={
  fill opacity=1,
  draw opacity=1,
  text opacity=1,
  at={(0,0)},
  anchor=south west,
  draw=white!80!black
},
log basis y={10},
tick align=outside,
tick pos=left,
x grid style={white!69.0196078431373!black},
xlabel={$P_{\text{PA}}$ [dBm]},
xmajorgrids,
xmin=12, xmax=32,
xminorgrids,
xtick style={color=black},
y grid style={white!69.0196078431373!black},
ylabel={SER},
ymajorgrids,
ymin=1e-04, ymax=1,
yminorgrids,
ymode=log,
ytick style={color=black}
]
\addplot [color5, line width=1.0pt, mark=|, mark size=2, mark options={solid}]
table {%
12.0577239990234 0.429658389091492
14.0517272949219 0.36518634557724
16.0287246704102 0.309902960062027
18.0006904602051 0.259627342224121
20.077751159668 0.222305911779404
22.0018272399902 0.193229824304581
24.0859298706055 0.171319884061813
26.0706672668457 0.152740699052811
28.0602493286133 0.144246906042099
30.032527923584 0.172317558526993
32.0257873535156 0.306855607032776
};
\addlegendentry{No DPD}
\addplot [color2, line width=1.0pt,  mark=o, mark size=2, mark options={solid}]
table {%
12.0541038513184 0.214992243051529
14.0394344329834 0.114347845315933
16.0387554168701 0.0422709763050079
18.0209884643555 0.0137694239616394
20.0337181091309 0.00458076000213623
22.043212890625 0.00169256925582886
24.0437927246094 0.000636667013168335
26.0692024230957 0.000256228446960449
28.0372848510742 0.000209647417068481
29.9942092895508 0.00185560584068298
31.7707443237305 0.16390141248703
};
\addlegendentry{TD-GMP~\cite{GMP_2006}}

\addplot [color3, line width=1.0pt,  mark=square, mark size=2, mark options={solid}]
table {%
12.1201283231336 0.233653023666970
14.2458003496763 0.125409074980643
16.0828068395673 0.0600304240813596
18.1429191620000 0.0207057088460451
20.1744678553296 0.00682208193027728
21.7682351323695 0.00301605920964092
24.0039273563572 0.000981241466140859
26.0614360224195 0.000460232078172891
28.0779738466701 0.000713495335180438
29.9652828109803 0.00230370742171078
31.9504499395424 0.0336545191855530
};
\addlegendentry{FD-NN~\cite{tarver2021virtual}}

\addplot [color4, line width=1.0pt,  mark=triangle, mark size=2, mark options={solid}]
table {%
12.08473777771 0.27754271030426
14.0860977172852 0.159118813276291
16.0300598144531 0.074243038892746
18.0574989318848 0.0281056106090546
20.040283203125 0.00939053893089294
22.0181884765625 0.00342951240539551
24.0206832885742 0.00128883719444275
26.0261344909668 0.00068325400352478
28.0062484741211 0.00102874040603638
30.0957679748535 0.00314443707466125
32.0510368347168 0.0477212905883789
};
\addlegendentry{FD-CNN}

\addplot [black, line width=1.0pt,  mark=asterisk, mark size=2, mark options={solid}]
table {%
12.048900604248 0.202166939973831
14.0170078277588 0.0971013391017914
16.0902328491211 0.0360365092754364
18.0346984863281 0.0115070086717606
20.070484161377 0.00326903998851776
22.073127746582 0.000963141322135925
24.037540435791 0.000363762378692627
26.0639381408691 0.000168879628181458
27.9365463256836 0.000127214193344116
29.9360656738281 0.000102084875106812
31.8653182983398 0.00527796149253845
};
\addlegendentry{Ideal PA~\cite{chani2018lower}}
\end{axis}
\end{tikzpicture}
	\vspace*{-0.5 \baselineskip}
	\caption{SER as a function of the average PA output power, $P_{\text{PA}}$.}
	\label{fig:SER_vs_P}
	\vspace*{-0.5 \baselineskip}
\end{figure}

Now we fix $B=32$ and $U=8$ for different DPD schemes from Fig.~\ref{fig:FLOPs_vs_B} and plot the \ac{ser} results as a function of the average PA output power, $P_{\text{PA}}$, in Fig.~\ref{fig:SER_vs_P}. Specifically, we plot the case of ideal linear-clipping PA~\cite{chani2018lower}. The linear-clipping PA has a linear behavior before the clipping region, which has the minimum distortion that any DPDs can achieve. The number of QAM symbols used to calculate the SER is around $10^6$.

We note that the SER curves exhibit two different behaviors depending on the nonlinear and clipping regions of the PAs. On one hand, as $P_{\text{PA}}$ is below $25$ dBm, most PAs in the massive MU-MIMO system operate in their linear/nonlinear regions, so the SER of all DPD cases improve dramatically with $P_{\text{PA}}$ since the \ac{snr} at the UEs are increasing accordingly. The proposed FD-CNN  requires $2.2\times$ and $8.9\times$ less complexity than TD-GMP and FD-NN, respectively, at the expense of some increase in the required power to achieve a certain SER. For example, it requires $0.8$ dBm and $1.8$ dBm more PA output power to achieve an SER $10^{-3}$. On the other hand, when $P_{\text{PA}}>25$ dBm, the clipping effect of the
PA brings unrecoverable distortions, which lead to a quick SER degradation for all DPD cases. The SER gap between all the DPD cases and the linear-clipping case may come from some residual distortions due to irreversible nonlinearity. 

\section{Conclusion}
We proposed a novel DPD method in the FD based on  CNNs to address the rising complexity issue of linearization for large antenna systems in massive MU-MIMO-OFDM. The complexity of the proposed FD-CNN DPD is in the magnitude of the number of UEs and the symbol rate, which naturally avoids high complexity 
as the number of BS antennas increases. Simulation results on a MU-MIMO-OFDM system with different behavior PAs on each RF chain show that the FD-CNN can save $2.2\times$ and $8.9\times$ number of FLOPs with $32$ BS antennas and $8$ UEs, compared with FD NN-based and TD per-antenna GMP-based DPDs, respectively. This saving can be further improved as the number of BS antennas increases. Furthermore, the SERs degradation of FD-CNN is minor. 

%\section*{Acknowledgment}
%This work was supported by the Swedish Foundation for Strategic Research (SSF), grant no. ID19-0021. The authors would like to thank Fan Jiang at Chalmers University of Technology for fruitful discussions.
%%%%%%%%%%%%%%%%%%%%%%%%%%%%%%%%%%%%%%%%%%%%%%%%%%%%%%%%%%%%%%%%%%%%%%%%%%%%%
\balance

\bibliographystyle{IEEEtran}

\bibliography{./bibliography/IEEEabrv,reference_list}
\end{document}